\documentclass[12pt,preprint]{aastex}  

\usepackage{epsfig}

\received{December 1, 2004}

\slugcomment{To appear in Astrophys. J.} 
\shorttitle{Heliosheath Interstellar Dust}
\shortauthors{Frisch et al.}

\def\OI{O$^{\rm o}$}
\def\HI{H$^{\rm o}$}
\def\HII{H$^{\rm +}$}
\def\HeI{He$^{\rm o}$}
\def\HeII{He$^{\rm +}$}
\def\nHI{$n$(H$^{\rm o}$)}
\def\nHII{$n$(H$^{\rm +}$)}
\def\nel{$n$(e$^{\rm -}$)}
\def\Bis{B$_{\rm IS}$}

\def\Cpb{\hbox{$C_{\rm P \beta}$}}
\def\gl{\hbox{$l$}}
\def\gb{\hbox{$b$}}
\def\el{\hbox{$\lambda$}}
\def\eb{\hbox{$\beta$}}

\def\kms{\hbox{km s$^{-1}$}}
\def\deeg{\hbox{$^{\rm o}$}}

\def\Lya{\hbox{Ly$\alpha$}}
\def\HeI{He$^{\rm o}$}

\def\cmtwo{cm$^{-2}$}
\def\thetac{$\theta_{\rm C}$}
\def\cc{cm$^{-3}$}

\begin{document}

\title{Tentative Identification of Interstellar Dust in Nose of the Heliosphere } 
\author{Priscilla C. Frisch}
\affil{Department of Astronomy and Astrophysics, University of
Chicago, Chicago, IL 60637.} \email{frisch@oddjob.uchicago.edu}

\begin{abstract}

Observations of the weak polarization of light from nearby stars,
reported by \citet{Tinbergen:1982}, are consistent with polarization 
by small, radius$<$0.14 $\mu$m, interstellar dust grains entrained in 
the magnetic wall of the heliosphere.  The region of maximum polarization 
is towards ecliptic coordinates ($\lambda$,$\beta$)$\sim$(295\deeg,0\deeg), 
corresponding to (\gl,\gb) = (20\deeg,--21\deeg), while the dust cone 
direction shows a marginally significant minimum in polarization.  
The direction of maximum polarization is offset along the ecliptic longitude 
by $\sim$35\deeg\ from the nose of the heliosphere, and extends to low 
ecliptic latitudes.  An offset is also seen between the region with the best
aligned dust grains, \el$\sim$281\deeg$\rightarrow$330\deeg, and the
upwind direction of the undeflected large grains, $\lambda \sim 259$\deeg, 
$\beta \sim +8$ \deeg, which are observed by Ulysses and Galileo to be 
flowing into the heliosphere.  In the aligned-grain region, the
strength of polarization anti-correlates with ecliptic latitude,
indicating that the magnetic wall is predominantly at negative ecliptic 
latitudes.  An extension of the magnetic wall to $\beta <$0\deeg\
is consistent with predictions by \citet{Linde:1998}.
A consistent interpretation follows if the maximum-polarization region
traces the heliosphere magnetic wall in a direction inclined to
the local interstellar magnetic field, \Bis, while the region of
best-aligned dust samples the region where \Bis\ stretchs smoothly
over the heliosphere with maximum compression.  These data are
consistent with a tilt of $\sim$60\deeg\ of \Bis\ with respect to the 
ecliptic plane, and parallel to the galactic plane.  Interstellar dust 
grains captured in the heliosheath may also introduce a weak, but important,
large scale contaminant for the cosmic microwave background signal
with a symmetry consistent with the relative tilts of \Bis\ and the
ecliptic.

\end{abstract}

\keywords{heliosphere --- interstellar : dust, interstellar --- dust: polarization}

\section{Introduction}

Tinbergen's detection of weak ($\ge 2 \sigma$) polarization of light
for $\sim$15 nearby stars, caused by magnetically aligned interstellar
dust grains in the galactic center hemisphere, has long been
intriguing \citep[][T82]{Tinbergen:1982}.  Instrumental noise may
introduce random weak polarization, and the significance of these
polarization data results partially from the consistent polarization
position angles for objects in a small region near the ecliptic plane
\citep[see Fig. 6 of T82 and Fig. 2 of][]{Frisch:2003jgr}.  T82 found the position angles to be consistent with
a galactic magnetic field directed toward $l \sim$70\deeg, which is
consistent with the $l \sim 80$\deeg\ local field direction found from
polarization measurements of more distant stars \citep{Heiles:1976p}.
The location of these magnetically aligned interstellar dust grains
(ISDGs) coincides with the nearest interstellar material (ISM) in the
galactic center hemisphere
\citep[e.g.][]{BruhweilerKondo:1982,FrischYork:1983}.

Previous discussions of the T82 data found little increase in
polarization with star distance, indicating the grains are close to
the Sun \citep{Frisch:1990,FrischSlavin:2005cospar}.  The regions 
of strongest polarization and most uniform
position angle are concentrated in the ecliptic plane near the
heliosphere nose region \citep{Frisch:2003jgr}.  Ulysses, Galileo, and
Cassini observe interstellar dust grains flowing into the heliosphere
from the heliosphere nose.  The grains capable of polarizing starlight,
radius $a> 0.05  ~ \mu$m, are part of the population of grains with 
$a<0.1 -0.2 ~ \mu$m that are filtered at the heliopause 
\citep[][L00,F99]{Baguhletal:1996,Linde:1998,Landgraf:2000,Frischetal:1999}.

In this note I argue that the T82 polarization data are consistent
with polarization by charged interstellar dust grains trapped in, and
diverted by, the interstellar magnetic field stretched over the
heliosphere.  Conventional grain alignment theories will need
evaluation for the unique outer heliosheath magneto-hydrodynamic (MHD)
configuration, where interstellar fields may be compressed by factors
of four or more, and field aligned currents are present
\citep[e.g.,][]{Lazarian:2000,Linde:1998,Ratkiewiczetal:1998,Pogorelovetal:2004}.
The T82 data, together with the 3 kHz signals detected by Voyager in
the outer heliosphere \citep{KurthGurnett:2003,Cairns:2004}, represent the primary
evidence for the interstellar magnetic field direction at the solar
location and are both consistent with a field direction towards $l
\sim$80\deeg.\footnote{After this paper was submitted, a discussion of
the interstellar magnetic field direction based on \Lya\
interplanetary glow data was presented by \citet{Lallementetal:2005}.}  If
my interpretation is correct, sensitive polarization observations over
the 22-year magnetic solar cycle should monitor the outer heliosheath
and detect variations in the interactions of these charged grains with
the heliosphere.

Small charged grains capable of polarizing starlight are magnetically
coupled to the interstellar magnetic field that is excluded from the
heliosphere.  For reasonable estimates of the far ultraviolet
radiation field, which causes photoelectric charging of the ISDGs, and
interstellar magnetic field strengths greater than 1.5 $\mu$G, dust
grains with radii $<$0.1--0.2 $\mu$m couple to the interstellar magnetic
field at the heliosphere.  ISDGs with radii $a > 0.05$ $\mu$m
provide most polarization in the diffuse ISM \citep{Mathis:2000}.
\citet{Linde:1998} estimates that ISDGs with radii $a<$0.14 $\mu$m
will not enter the heliosphere because of deflection through large
angles in a magnetic wall, which is formed by compressed interstellar fields
lines stretched over the heliosphere nose.  For the case where
\Bis$\sim$1.5 $\mu$m, and a field angle of 60\deeg\ with respect to the
ecliptic plane, field strength increases by factors of $\sim$4--5, or
more, in the resulting magnetic wall.  However, an even stronger \Bis\ is
expected if magnetic and thermal energies are approximately equal in the ISM
surrounding the heliosphere.  For thermal energy density $E_{\rm th}/k = 1.5 nT \sim
3700$ \cc\ K, magnetic energy density $E_{\rm B}= $\Bis$^2$/$ 8 k \pi$ \cc\
K, and $E_{\rm th} \sim E_{\rm B}$, then \Bis$\sim$3.6 $\mu$G for the
local ISM, which has temperature and densities T=6340 K, \nHI$\sim$0.20 \cc,
\nHII$\sim$0.09 \cc, and \nel$\sim$0.1 \cc\
\citep{SlavinFrisch:2002,FrischSlavin:2003}.

Larger ISDGs flow into the heliosphere at $\sim$26.3 \kms\ ($\sim$5
AU/year) from the upwind direction, $\lambda$=259\deeg,
$\beta$=+8\deeg, and have been measured at all ecliptic latitudes in
the inner 5 AU of the heliosphere
\citep[e.g.,][]{Baguhletal:1996,Landgraf:2000,CzechowskiMann:2003}.
ISDG trajectories depend on the charge-to-mass ratio and polarity of
the solar magnetic cycle.  
Intermediate-sized charged grains, $a \sim $0.2 $\mu$m, couple to the
solar wind by the Lorentz force, and are alternately focused and
defocussed by the changing polarity of the solar-cycle.  For the
positively charged ISDGs, these periods coincided with grain
defocusing cycles (L00).  Large grains ($a >$0.5 $\mu$m) are
gravitationally focused downwind of the Sun leaving a trail (or
``focusing cone'') of interstellar dust extending for $>$10 AU.  A
similar focusing cone is seen in interstellar \HeI\ data
\citep{Witte:2004,Moebiusetal:2004}.

The magnetic polarity of the Sun was North-positive during
$\sim$1971.6--1980.2 when the T82 data are likely to have been
acquired, and again when U/G and Voyager 3 kHz data were acquired in
the 1990s \citep[e.g.][]{Frischetal:2005}.

\section{Tinbergen Polarization Data}

Tinbergen observed $\sim$180 stars at 1$\sigma$ levels of degree of
polarization of 7 x 10$^{-5}$, and concluded that there is a region of
interstellar dust creating weak polarization of the light from nearby
($<$40 pc) stars, with the dust centered around the galactic interval
of \gl$\sim$340\deeg$\pm$40\deeg, \gb$\sim$0\deeg.  This direction is
consistent with the LSR\footnote{Here, LSR is the Local Standard of
Rest as defined by the Standard solar apex motion.}  direction of
upwind flow for the cluster of local interstellar clouds (CLIC),
\gl=331.4\deeg, \gb=--4.9\deeg\ \citep[and velocity $V \sim$--19.4
\kms,][]{FGW:2002}.  (The corresponding position in ecliptic
coordinates is $\el$=255.6\deeg and $\eb$=--32.6\deeg.)  The
discussion here is restricted to $\sim$160 stars within 40 pc of the
Sun.  T82 reported Stokes parameters $Q$ and $U$ for three channels,
I, II, III, based on filters centered near 5400, 6100, and 8000 A,
respectively, along with the averages of channels I and II.  The
channel I, II averages for $Q$, $U$ are used in this paper (as listed
in columns 10 and 11, Table 5, of T82).  Polarization is given by
$P=(Q^2 + U^2)^{1/2}$, and position angle in celestial coordinates
\thetac=0.5 arctan($U$/$Q$) \citep[also see][]{Heiles:2000}.  Note the
standard convention is used for defining the angle of polarization,
\thetac, so that $Q$ is positive and $U$ is zero when the electric
vector is North-South in the equatorial (celestial) coordinate system.
It is shown below that these Tinbergen data show a distinct signature
which is related to the ecliptic geometry, and which is consistent
with a magnetic wall in the heliosphere nose.

The T82 observations of this nearby region of enhanced weak
polarization were not reproduced by a survey of $\sim$400 stars, with
an accuracy of $\sigma \sim$2 10$^{-4}$ \citep[][L93]{Leroy:1993a}.
Eleven stars in the T82 patch were observed in both surveys, with
declinations down to --30\deeg.  
Although Leroy does not confirm the T82 results, his data are
not inconsistent with the T82 results.
The observation dates are
unclear in the original T82 and L93 papers, but it appears the T82
data were acquired in the years surrounding or following the 1975 
solar minimum, whereas the L93 data
were acquired near the 1990--1992 solar maximum, where the outer
heliosheath configuration would have been different.

The strength of the anticorrelation between polarization and ecliptic
latitude is shown in Fig. \ref{fig:1}.  It has been determined from the
covariance, \Cpb, of polarization $P$ and ecliptic latitude \eb,
where:
\begin{equation}
 C_{\rm P \beta}  = \frac{1}{\rm (N-1)*(P_{\rm var}*\beta_{\rm var})^{0.5}}\sum_{\rm i}^{\rm N}(P_{\rm
i}-\overline{P})*(\beta_{\rm i}-\overline{\beta}).  
\end{equation}
Here $\overline{P}$ and $P_{\rm var}$ (and $\overline{\beta}$ and
$\beta_{\rm var}$) are the mean and variance of $P$ (and $\beta$)
respectively, calculated for stars in the interval $\lambda_0
\pm$20\deeg\ centered at an arbitrary ecliptic longitude
$\lambda_0$, while N is the number of stars in that interval.  N
ranges from 5 to 17 for the points plotted in Fig. \ref{fig:1}.  A covariance
factor of $\sim$--0.5 is found for stars near the upwind direction,
but in addition offsets between inflowing and polarizing dust grains
reveal details about the interaction of the heliosphere and
interstellar magnetic field (see below).  The significance of this
covariance is tested by performing a similar analysis, using the same
star sample, but with different assumptions.  In the first case, an
equivalent calculation of \Cpb\ is completed using galactic instead of
ecliptic coordinates, and the only features appearing simultaneously
(for the same $\lambda_0$) in \Cpb, $P$, and \thetac\ are towards the
galactic center (corresponding to the heliosphere nose direction), and
a group of polarized stars near (\gl,\gb)$\sim$(55\deeg,+55\deeg).
For the second test, a set of values were generated for the Stokes
parameters, Q and U, with values randomly distributed between 0 and 35
(corresponding to a 5 $\sigma$ polarization, $P=35 \times 10^{-5}$ deg. polarization).  These random polarizations
were subjected to the same analysis as the real data, and the
equivalent of Fig. \ref{fig:1} is essentially a scatter plot for $P$, \thetac,
and \Cpb, with no coherent patterns that depend on $\lambda$.  A third
test uses the 25 stars contributing to the points in the interval
$\lambda_0 =280-325$\deeg, which dominate the observed
anticorrelation.  The polarization for 13 stars with $\beta < 0$ is $
P = 16.3 \pm 6.4 \times 10^{-5}$ deg, while the average polarization
for the 12 stars with $\beta > 0$ is $ P = 8.0 \pm 6.1 \times 10^{-5}$
deg.  Applying the Student t-test to these two samples gives an
estimate, at the 98\% confidence level, that the polarizations for these
two samples are not drawn from a single sample with randomly
distributed polarizations.  Thus, the anticorrelation between
polarization and $\beta$, for $\lambda = 280-325$\deeg, appears real.
This anticorrelation indicates that the polarization signal is 
dominated by stars with $\beta<$0\deeg\ in this interval.

Fig. \ref{fig:1} summarizes the properties of the polarization data
for stars within 50\deeg\ of the ecliptic plane. Each plotted point
represents properties averaged over a 40\deeg\ longitude interval,
centered at an ecliptic longitude $\lambda_0$.  The longitude,
$\lambda_0$, is stepped along the ecliptic plane at intervals of
3\deeg\ in order to display variations that depend on ecliptic
longitude.  Fig. \ref{fig:1}, top, shows the variation in polarization, $P$, as
a function of $\lambda_0$.  Fig. \ref{fig:1}, middle, shows that an interval
extending from $\lambda_0$=281\deeg$\rightarrow$330\deeg\ exhibits
highly aligned grains (where \thetac$\sim -35$\deeg, for the
polarization angle in celestial coordinates), and encompasses the
direction of maximum polarization observed towards
\el$\sim$295\deeg.\footnote{The position of (\el,\eb) =
(295\deeg,0\deeg) corresponds to (\gl,\gb) = (20\deeg,--21\deeg).}  In
the interval showing the strongest and most consistent polarization
angle (\el=281\deeg$\rightarrow$330\deeg), the correlation coefficient
between $P$ and \eb\ is $C_{\rm P \beta} \sim$--0.5 (bottom panel,
Fig. \ref{fig:1}).  The strength of the $P$---\eb\ anticorrelation for only
those stars with $\eb <$0\deeg\ gives $C_{\rm P \beta} \sim$--0.7, a maximum
smoothed value of $P \sim 20 \times 10^{-5}$ degrees, and a direction of
maximum $P$ towards \el$\sim$294\deeg$\pm$4\deeg.  The anomalously
high values of $C_{\rm P \beta} \sim$+0.5, found at \el$\sim$60\deeg,
are dominated by the two non-variable stars HD 38393 (F7 V, 9 pc) and
HD 40136 (F1 V, 15 pc), located near (\gl,\gb) $\sim$
(223\deeg,--22.\deeg), and with $P \sim$15 10$^{-5}$ degrees.  The
central direction of the \HeI\ cone (Fig. \ref{fig:1}, top) corresponds to a
minimum in the polarization strength, and it extends for
$\sim$15\deeg.  Since grains in the dust cone are larger than typical
grains which polarize optical light \citep{Landgraf:2000}, this
minimum, although marginally significant, could be explained if real.
The inflowing dust grains observed by Ulysses and Galileo (U/G,
Fig. \ref{fig:1}, top) tend to be larger than grains captured in the heliosheath
(F99), and have a best-fit upwind direction within $\sim$5\deeg\ of
the upwind direction, as defined by \HeI\ data.
Fig. \ref{fig:1} shows clearly that the region of maximum dust inflow is offset
from the region of maximum dust alignment.

In contrast, if one assumes a purely interstellar origin for the
polarization, and applies standard ISM values ($P/A_{\rm V}<$0.03,
$A_{\rm V}/E(B-V)$=3.1, and $N$(H)/$E(B_V)$=5.8 x 10$^{21}$ \cmtwo),
then a 1$\sigma$ polarization corresponds to a cloud column density of
$N$(H)$\sim$4 x 10$^{18}$ \cmtwo\ for a magnetic field perpendicular
to the sightline.  This value is consistent with expected amount of
nearby upwind interstellar gas; for instance, toward 36 Oph
$N$(HI)=7.1 10$^{17}$ \cmtwo \citep{Wood36Oph:2000}, and the gas in
this sightline may be partially ionized \citep{Frisch:2004}.  This
argument, in turn, implies a purely interstellar origin for the
polarization, with a possible small polarization enhancement in
heliosheath currents.  However, in this case it is difficult to
explain the ecliptic signatures on starlight polarization as shown in
Figs. 1 and 2.  The gas-to-dust mass ratio is, in any case, uncertain
for such small reddening values.

\section{Discussion}

ISDG interactions with the outer heliosheath may depend on solar cycle
phase.  During solar minimum phases, the heliospheric HI \Lya\ glow
should show a pronounced groove from the asymmetric momentum flux of
the solar wind \citep{Bzowski:2003}, compared to the more symmetric
(although smaller overall) heliosphere during solar maximum.  These
differences in heliospheric morphology will affect the interstellar
magnetic field and dust interactions with the heliosphere, and may
explain the lack of confirmation of the T82 data by L93.

For magnetically aligned ISDGs in space, the plane of polarization is
parallel to \Bis, and maximum polarization will be seen for directions
perpendicular to the field lines \citep{Heiles:1976p}.  The
polarization maximum is offset by $\sim+$30\deeg$\pm$5\deeg\ from the
heliosphere nose, and should trace thick regions of the magnetic wall
where the sightline is relatively perpendicular to the field
direction.  The direction of \Bis\ is \gl$\sim$70--80\deeg\
\citep[from Tinbergen and][]{Heiles:1976p}, which indicates that \Bis\
is tilted by $\sim$60\deeg\ with respect to the ecliptic plane.
The region of maximum polarization is centered near \el$\sim$295\deeg, but several
strongly polarized stars are seen at low latitudes between
(\el,\eb)$\sim$(280\deeg,--10\deeg) and (320\deeg--40\deeg).  The
strong polarization in this region at low ecliptic latitudes, $\beta
\sim$--40\deeg, may originate in the low latitude extension of the magnetic
wall resulting from the tilt of \Bis\ with respect to the ecliptic
plane.  Linde (1998) modeled the magnetic wall for the 1996 solar
minimum, and found it stronger at southern latitudes where the
azimuthal components of the interstellar and interplanetary fields are
parallel, as compared to the northern hemisphere where they were
antiparallel.  The best aligned grains (\el=281\deeg--330\deeg) should
trace compressed \Bis\ field lines which stretch smoothly around the
heliosphere \citep{Linde:1998,Pogorelovetal:2004,Ratkiewiczetal:1998}.
Although the alignment mechanism is somewhat uncertain, polarization
may be enhanced in the heliosheath nose by grain charging (e.g. by the
Barnett effect) and the tight coupling between the interstellar dust
grains and the compressed \Bis\ \citep[see the discussion of grain
alignment in][]{Lazarian:2000}.  However, detailed models of heliosphere grain
alignment and trapping are required before these results can be fully
understood.

In Fig. \ref{fig:2}, the distribution of polarization strengths are plotted 
in ecliptic coordinates, together with the upwind direction of
the interstellar dust, \HI, and \HeII\ flows.
The positions of the 3 kHz bursts are also plotted.  
Both the distribution of 3 kHz bursts and the alignment
of the polarization directions, shown in Fig. 2 of Frisch, 2003
and Fig. 6. of T82, indicate that \Bis\ is relatively parallel to the 
galactic plane.  This orientation corresponds to a tilt by $\sim$60\deeg\ 
with respect to the ecliptic plane.  
The inflowing dust grains observed by Ulysses and Galileo (U/G, Fig. \ref{fig:1}, top) 
tend to be larger than grains captured in the heliosheath (F99), and have a
best-fit upwind direction within $\sim$5\deeg\ of the upwind direction
(as defined by the antipode of the \HeI\ cone).  However the 2$\sigma$
uncertainties on the U/G flow direction extend to smaller \el\ values,
corresponding to \el=210\deeg$\rightarrow$285\deeg.  Fig. \ref{fig:1} shows
clearly that the region of maximum dust inflow (large grains) is
offset from the region of maximum dust alignment (deflected small
grains).  Fig. \ref{fig:2} shows that the direction of maximum polarization,
which should trace the magnetic wall, is inclined by a large angle to
the ecliptic plane.  This offset between aligned and inflowing grains
also indicates that dust filtration reflects the asymmetric
heliosphere configuration caused by \Bis, with the large-grain inflow
showing the heliosphere nose, and the small grains showing the
magnetic configuration of the outer heliosheath.

In principle, the relative distributions of the aligned dust, dust
inflow, and \HeI\ and \HI\ upwind directions (see Fig. \ref{fig:2})
will be understood if we impose the requirement that the filtration
factors for dust, \HII, and other charged species vary with their
gyroradius in the magnetic wall.  Small dust grains are excluded
(radii less than $\sim$0.05--0.1 $\mu$m) and cause maximum
polarization in directions parallel to \Bis.  Large grains (radii
$>>$ 0.2 $\mu$m) experience minimal filtration.  About 50\% of the
\HI\ is filtered in the outer heliosheath.  Protons are initially
deflected perpendicular to \Bis, but become diverted around the
heliosphere along with \Bis\ in the magnetic wall.  In Fig. \ref{fig:2}, the
stars with the strongest polarization form a band which makes an angle
of $\sim$65\deeg\ with respect to the ecliptic plane, and similar
angles are seen between the offsets of the \HI\ and \HeI\ upwind
directions.  It seems a good guess that this alignment traces the
magnetic wall orientation caused by the distortion of \Bis\ at the
heliosphere.  The IBEX data on fast \HI\ and \OI\ neutral atoms formed
in the heliosheath \citep{McComasetal:2004} may map out this
heliosphere asymmetry driven by the interstellar magnetic field,
through observations of \HI\ and \OI\ fast neutrals, which have
formation rates that depend on filtration factors.

Future precise observations of very weak polarization signals, with
duplicate observations and using rotatable telescopes in the northern
and southern hemispheres, may provide a useful monitor of the outer
heliosheath region, and of the interaction between the solar and
interstellar magnetic fields.  Detailed models of heliosphere grain
alignment and trapping are required before these results can be fully
understood.

Removing contributions from foreground emission is an important
element in analyzing WMAP composite maps \citep{Bennett:2003}.  The
possibility of a weak large scale contributions from the heliosphere
indicates further modeling of this emission is warranted
\citep{FrischHanson:2004,FrischSlavin:2004}.  The infrared emission
from heliospheric interstellar dust appears much weaker than zodiacal
emission, by factors of $\sim$10$^2$ (F99).  However, the observed
correlation with the ecliptic of the combined quadrupole-octopole signature
found by \citet{Schwarzetal:2004} in the WMAP data, supports a possible
contamination from ISDGs interacting with the heliosphere.  Candidates
for contamination include the small polarizing grains 
trapped in the magnetic wall, and discussed here; current sheets in the outer
heliosheath regions; or alternatively from larger heated interstellar
dust interacting with the solar wind.  Any contribution to the
cosmic microwave background from small grains in the outer
heliosheath regions should reflect the complex asymmetry of the
heliosphere interacting with \Bis, including the magnetic wall,
rather than echoing the more simple
symmetry of the ecliptic plane.  If the smaller grains, radii $a<$0.2
$\mu$m, are responsible, the spatial distribution may show a variation
with the solar cycle, such that the heliospheric contribution to the
cosmic microwave signal could be recovered from sensitive polarization
observations spaced throughout the 22 year magnetic solar cycle.


\section{Acknowledgements}
The author would like to thank
NASA (grants NAG5-11005, NAG5-13107) for supporting this work.

\begin{thebibliography}{35}
\expandafter\ifx\csname natexlab\endcsname\relax\def\natexlab#1{#1}\fi

\bibitem[{{Baguhl} {et~al.}(1996){Baguhl}, {Gruen}, \&
  {Landgraf}}]{Baguhletal:1996}
{Baguhl}, M., {Gruen}, E., \& {Landgraf}, M. 1996, Space Science Reviews, 78,
  165

\bibitem[{{Bennett} {et~al.}(2003){Bennett}, {Hill}, {Hinshaw}, {Nolta},
  {Odegard}, {Page}, {Spergel}, {Weiland}, {Wright}, {Halpern}, {Jarosik},
  {Kogut}, {Limon}, {Meyer}, {Tucker}, \& {Wollack}}]{Bennett:2003}
{Bennett}, C.~L., {Hill}, R.~S., {Hinshaw}, G., {Nolta}, M.~R., {Odegard}, N.,
  {Page}, L., {Spergel}, D.~N., {Weiland}, J.~L., {Wright}, E.~L., {Halpern},
  M., {Jarosik}, N., {Kogut}, A., {Limon}, M., {Meyer}, S.~S., {Tucker}, G.~S.,
  \& {Wollack}, E. 2003, \apjs, 148, 97

\bibitem[{{Bruhweiler} \& {Kondo}(1982)}]{BruhweilerKondo:1982}
{Bruhweiler}, F.~C. \& {Kondo}, Y. 1982, \apj, 259, 232

\bibitem[{{Bzowski}(2003)}]{Bzowski:2003}
{Bzowski}, M. 2003, \aap, 408, 1155

\bibitem[{{Cairns}(2004)}]{Cairns:2004}
{Cairns}, I.~H. 2004, in AIP Conf. Proc. 719: Physics of the Outer Heliosphere,
  381--386

\bibitem[{{Czechowski} \& {Mann}(2003)}]{CzechowskiMann:2003}
{Czechowski}, A. \& {Mann}, I. 2003, \aap, 410, 165

\bibitem[{{Frisch}(1990)}]{Frisch:1990}
{Frisch}, P.~C. 1990, in Physics of the Outer Heliosphere, ed. S.~Grzedzielski
  \& D.~E. Page, 19--22

\bibitem[{{Frisch}(2003)}]{Frisch:2003jgr}
{Frisch}, P.~C. 2003, \jgr, 108, 11

\bibitem[{{Frisch}(2004)}]{Frisch:2004}
{Frisch}, P.~C. 2004, in AIP Conf. Proc. 719: Physics of the Outer Heliosphere,
  404--411

\bibitem[{{Frisch} {et~al.}(1999){Frisch}, {Dorschner}, {Geiss}, {Greenberg},
  {Gr\"un}, {Landgraf}, {Hoppe}, {Jones}, {Kr{\"{a}}tschmer}, {Linde},
  {Morfill}, {Reach}, {Slavin}, {Svestka}, {Witt}, \& {Zank}}]{Frischetal:1999}
{Frisch}, P.~C., {Dorschner}, J.~M., {Geiss}, J., {Greenberg}, J.~M., {Gr\"un},
  E., {Landgraf}, M., {Hoppe}, P., {Jones}, A.~P., {Kr{\"{a}}tschmer}, W.,
  {Linde}, T.~J., {Morfill}, G.~E., {Reach}, W., {Slavin}, J.~D., {Svestka},
  J., {Witt}, A.~N., \& {Zank}, G.~P. 1999, \apj, 525, 492

\bibitem[{{Frisch} {et~al.}(2002){Frisch}, {Grodnicki}, \& {Welty}}]{FGW:2002}
{Frisch}, P.~C., {Grodnicki}, L., \& {Welty}, D.~E. 2002, \apj, 574, 834

\bibitem[{{Frisch} \& Hanson(2004)}]{FrischHanson:2004}
{Frisch}, P.~C. \& Hanson, A.~J. 2004, unpublished

\bibitem[{{Frisch} {et~al.}(2005){Frisch}, {M{\" u}ller}, {Zank}, \&
  {Lopate}}]{Frischetal:2005}
{Frisch}, P.~C., {M{\" u}ller}, H.~R., {Zank}, G.~P., \& {Lopate}, C. 2005, in
  Astrophysics of Life, 21--34

\bibitem[{{Frisch} \& {Slavin}(2003)}]{FrischSlavin:2003}
{Frisch}, P.~C. \& {Slavin}, J.~D. 2003, \apj, 594, 844

\bibitem[{{Frisch} \& Slavin(2004)}]{FrischSlavin:2004}
{Frisch}, P.~C. \& Slavin, J.~D. 2004, \ssr, 0000

\bibitem[{{Frisch} \& {Slavin}(2005)}]{FrischSlavin:2005cospar}
{Frisch}, P.~C. \& {Slavin}, J.~D. 2005, Adv.Sp. Res., in press

\bibitem[{{Frisch} \& {York}(1983)}]{FrischYork:1983}
{Frisch}, P.~C. \& {York}, D.~G. 1983, \apjl, 271, L59

\bibitem[{{Heiles}(1976)}]{Heiles:1976p}
{Heiles}, C. 1976, \araa, 14, 1

\bibitem[{{Heiles}(2000)}]{Heiles:2000}
---. 2000, \aj, 119, 923

\bibitem[{{Kurth} \& {Gurnett}(2003)}]{KurthGurnett:2003}
{Kurth}, W.~S. \& {Gurnett}, D.~A. 2003, \jgr, 108, 2

\bibitem[{{Lallement} {et~al.}(2005){Lallement}, {Qu{\' e}merais}, {Bertaux},
  {Ferron}, {Koutroumpa}, \& {Pellinen}}]{Lallementetal:2005}
{Lallement}, R., {Qu{\' e}merais}, E., {Bertaux}, J.~L., {Ferron}, S.,
  {Koutroumpa}, D., \& {Pellinen}, R. 2005, Science, 307, 1447

\bibitem[{{Landgraf}(2000)}]{Landgraf:2000}
{Landgraf}, M. 2000, \jgr, 105, 10303

\bibitem[{{Lazarian}(2000)}]{Lazarian:2000}
{Lazarian}, A. 2000, in ASP Conf. Ser. 215, 69

\bibitem[{{Leroy}(1993)}]{Leroy:1993a}
{Leroy}, J.~L. 1993, \aap, 274, 203

\bibitem[{Linde(1998)}]{Linde:1998}
Linde, T.~J. 1998, PhD thesis, Univ. of Michigan, Ann Arbor, {small \tt
  http://hpcc.engin.umich.edu/CFD/publications}

\bibitem[{{M{\" o}bius} {et~al.}(2004){M{\" o}bius}, {Bzowski}, {Chalov},
  {Fahr}, {Gloeckler}, {Izmodenov}, {Kallenbach}, {Lallement}, {McMullin},
  {Noda}, {Oka}, {Pauluhn}, {Raymond}, {Ruci{\' n}ski}, {Skoug}, {Terasawa},
  {Thompson}, {Vallerga}, {von Steiger}, \& {Witte}}]{Moebiusetal:2004}
{M{\" o}bius}, E., {Bzowski}, M., {Chalov}, S., {Fahr}, H.-J., {Gloeckler}, G.,
  {Izmodenov}, V., {Kallenbach}, R., {Lallement}, R., {McMullin}, D., {Noda},
  H., {Oka}, M., {Pauluhn}, A., {Raymond}, J., {Ruci{\' n}ski}, D., {Skoug},
  R., {Terasawa}, T., {Thompson}, W., {Vallerga}, J., {von Steiger}, R., \&
  {Witte}, M. 2004, \aap, 426, 897

\bibitem[{{Mathis}(2000)}]{Mathis:2000}
{Mathis}, J.~S. 2000, \jgr, 10269

\bibitem[{{McComas} {et~al.}(2004){McComas}, {Allegrini}, {Bochsler},
  {Bzowski}, \~{Collier}, {Fahr}, {Fichtner}, {Frisch}, \~{Funsten},
  {Fuselier}, {Gloeckler}, {Gruntman}, \~{Izmodenov}, {Knappenberger}, {Lee},
  {Livi}, \~{Mitchell}, {M{\" o}bius}, {Moore}, {Reisenfeld}, \~{Roelof},
  {Schwadron}, {Wieser}, {Witte}, \~{Wurz}, \& {Zank}}]{McComasetal:2004}
{McComas}, D., {Allegrini}, F., {Bochsler}, P., {Bzowski}, M., \~{Collier}, M.,
  {Fahr}, H., {Fichtner}, H., {Frisch}, P., \~{Funsten}, H., {Fuselier}, S.,
  {Gloeckler}, G., {Gruntman}, M., \~{Izmodenov}, V., {Knappenberger}, P.,
  {Lee}, M., {Livi}, S., \~{Mitchell}, D., {M{\" o}bius}, E., {Moore}, T.,
  {Reisenfeld}, D., \~{Roelof}, E., {Schwadron}, N., {Wieser}, M., {Witte}, M.,
  \~{Wurz}, P., \& {Zank}, G. 2004, in AIP Conf. Proc. 719: Physics of the
  Outer Heliosphere, 162--181

\bibitem[{{Pogorelov} {et~al.}(2004){Pogorelov}, {Zank}, \&
  {Ogino}}]{Pogorelovetal:2004}
{Pogorelov}, N.~V., {Zank}, G.~P., \& {Ogino}, T. 2004, \apj, 614, 1007

\bibitem[{{Ratkiewicz} {et~al.}(1998){Ratkiewicz}, {Barnes}, {Molvik},
  {Spreiter}, {Stahara}, {Vinokur}, \& {Venkateswaran}}]{Ratkiewiczetal:1998}
{Ratkiewicz}, R., {Barnes}, A., {Molvik}, G.~A., {Spreiter}, J.~R., {Stahara},
  S.~S., {Vinokur}, M., \& {Venkateswaran}, S. 1998, \aap, 335, 363

\bibitem[{{Schwarz} {et~al.}(2004){Schwarz}, {Starkman}, {Huterer}, \&
  {Copi}}]{Schwarzetal:2004}
{Schwarz}, D.~J., {Starkman}, G.~D., {Huterer}, D., \& {Copi}, C.~J. 2004,
  Phys. Rev. Let., 93, 221301

\bibitem[{{Slavin} \& {Frisch}(2002)}]{SlavinFrisch:2002}
{Slavin}, J.~D. \& {Frisch}, P.~C. 2002, \apj, 565, 364

\bibitem[{Tinbergen(1982)}]{Tinbergen:1982}
Tinbergen, J. 1982, \aap, 105, 53

\bibitem[{{Witte}(2004)}]{Witte:2004}
{Witte}, M. 2004, \aap, 426, 835

\bibitem[{{Wood} {et~al.}(2000){Wood}, {Linsky}, \& {Zank}}]{Wood36Oph:2000}
{Wood}, B.~E., {Linsky}, J.~L., \& {Zank}, G.~P. 2000, \apj, 537, 304

\end{thebibliography}

\begin{figure} 
\epsscale{.60}
\plotone{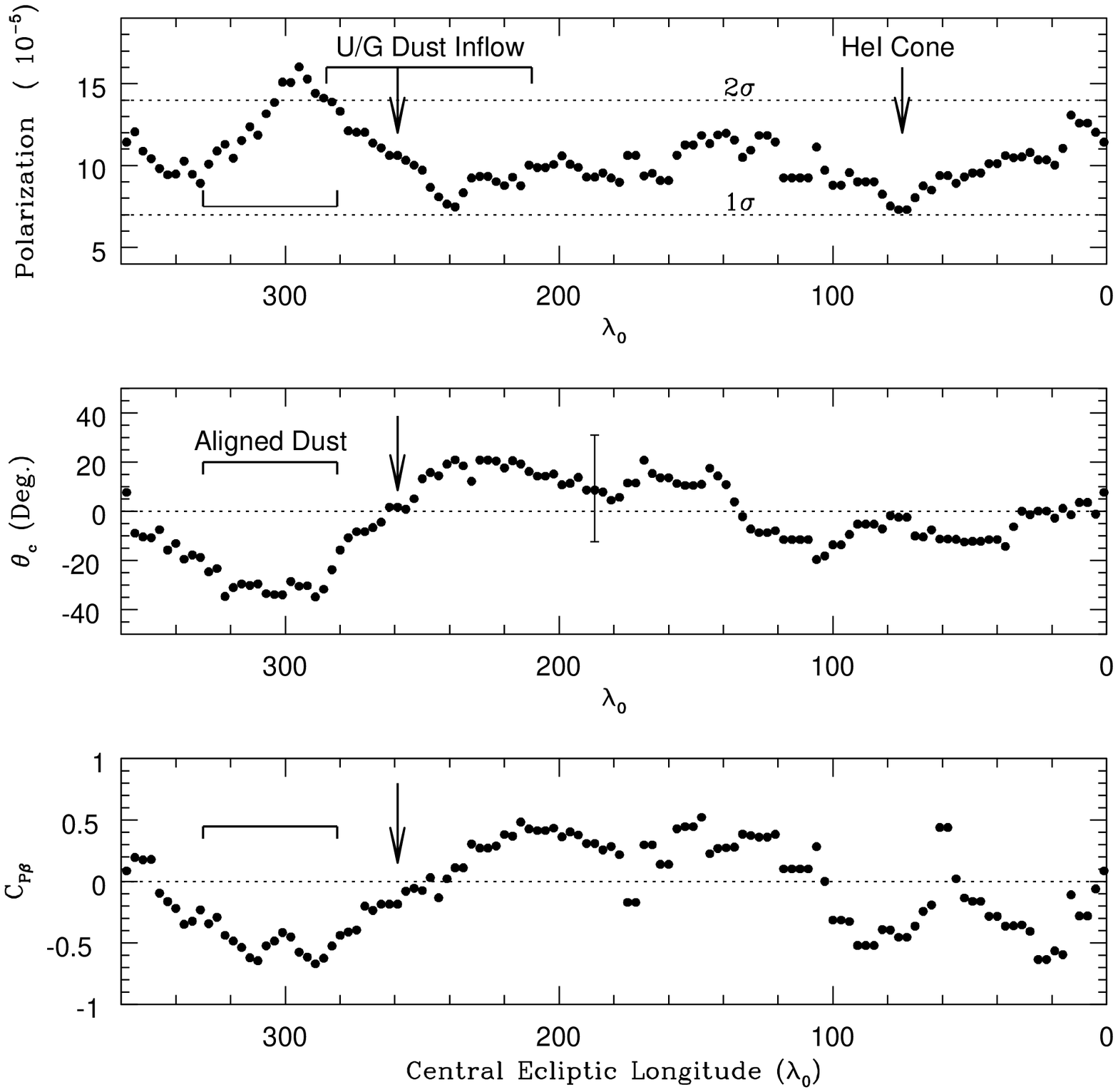} 
\end{figure}
\newpage
\begin{figure} 
\caption{
\label{fig:1} 
Various properties are displayed for the polarization data towards
stars within 50\deeg\ of the ecliptic plane.  The polarization
properties of stars within $\pm$20\deeg\ of a given ecliptic longitude
$\lambda_o$ are averaged together.  The displayed points represent
3\deeg\ increments in $\lambda_o$, as the direction sweeps from
$\lambda$=0\deeg\ (right) to $\lambda$=360\deeg\ (left).  Various
properties of the dust polarization are clearly related to position in
the ecliptic plane, such as region of maximum polarization and the
angle of polarization.  Top panel: The degree of polarization is
shown, together with 1$\sigma$ and 2$\sigma$ uncertainties in the
degree of polarization as quoted by Tinbergen.  The $\lambda$
direction and 2$\sigma$ uncertainties of the best fitting inflow
direction as determined from the Ulysses and Galileo observations of
interstellar dust inside of the solar system are shown as an arrow and
bar (respectively, from Fig. 9 in F99).  The upwind gas and dust
directions differ by $\sim$5\deeg.  The central direction of the \HeI\
cone (in the downwind direction) is plotted
\citep[from][]{Witte:2004}.  Middle: The polarization angle, \thetac\
(given in the original celestial coordinates of T82), is shown for the
same set of stars.  The region of
$\lambda$=281\deeg$\rightarrow$330\deeg\ shows a consistent angle of
polarization, where the dust grains have their maximum alignment.  The
error bar shows 1$\sigma$ uncertainties on \thetac.  Bottom: The
correlation coefficient between degree of polarization (top panel) and
ecliptic latitude is plotted as a function of the ecliptic longitude.
}
\end{figure}

\clearpage 

\begin{figure} 
\epsscale{.50}
\plotone{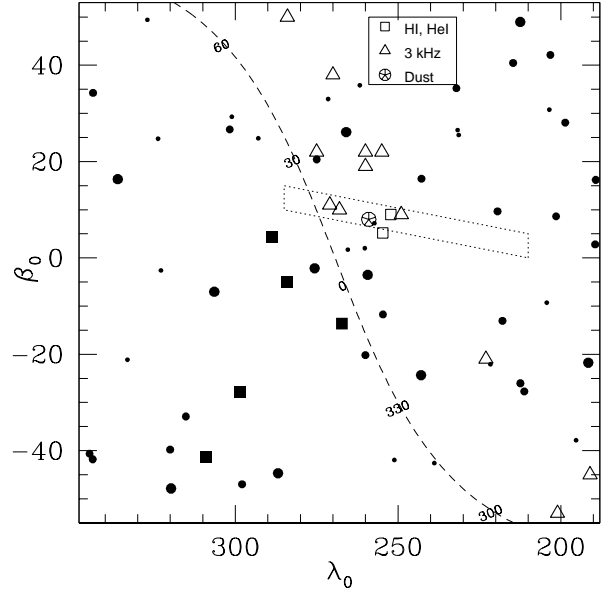} 
\caption{
\label{fig:2} 
Polarization strengths are plotted in ecliptic coordinates, for stars
within 40 pc and near the heliosphere nose direction.  The filled
circles represent polarizations of $P<3 \sigma$, and squares show $P>3
\sigma$.  Small, medium and large circles are for $P<1 \sigma$, $1
\sigma - 2 \sigma$ and $2 \sigma- 3 \sigma$.  Also plotted are the
locations of 3 kHz emission bursts signals detected by Voyager
\citep{KurthGurnett:2003}, the \HeI\ and \HI\ upwind directions
\citep{Witte:2004,Lallementetal:2005}, the inflow direction of
interstellar dust as measured by Ulysses and Galileo (circled star)
and an approximation of the U/G 2$\sigma$ error box \citep[dotted
lines,][]{Frischetal:1999}.  The galactic plane is shown as a dashed
line.  The region of maximum polarization (squares) appears to
indicate the magnetic wall caused by maximum compression of
interstellar \Bis\ stretched over the heliosphere.  The polarization
angle (not plotted) of these magnetically aligned ISDGs indicates that
\Bis\ is approximately parallel to galactic plane and and inclined to
the ecliptic by $\sim$60\deeg, while the direction of maximum
polarization traces the magnetic wall.  }
\end{figure}

\end{document}